\documentclass[aps,pra,showpacs,twocolumn,nofootinbib]{revtex4-2}
\usepackage[colorlinks=true, allcolors=blue]{hyperref}
\usepackage{bm}
\usepackage{amsmath}\allowdisplaybreaks
\usepackage{amssymb}
\usepackage{slashed}
\usepackage{float}
\usepackage{relsize}

\usepackage{subcaption}
\usepackage{dcolumn}
\usepackage{graphicx}
\usepackage{xcolor}
\usepackage{nicefrac}
\usepackage{physics}
\usepackage{multirow}
\usepackage{booktabs}

\newcolumntype{.}{D{x}{}{-1}}

\newcolumntype{w}[1]{D{.}{.}{#1}}
\newcolumntype{L}{>{$}l<{$}}
\newcommand{\balpha}{\vec{\alpha}}

\newcommand{\bfp}{\vec{p}}

\newcommand{\bfr}{\vec{r}}

\newcommand{\Za}{{Z\alpha}}

\newcommand{\lbr}{\langle}
\newcommand{\rbr}{\rangle}

\newcommand{\vare}{\varepsilon}

\begin{document}

\title{Higher-order corrections to the field shift in atomic systems}

\author{V.~A. Yerokhin}
\email{ vladimir.yerokhin@mpi-hd.mpg.de}
\affiliation{Max~Planck~Institute for Nuclear Physics, Saupfercheckweg~1, D~69117 Heidelberg, Germany}

\author{Z. Harman}
\affiliation{Max~Planck~Institute for Nuclear Physics, Saupfercheckweg~1, D~69117 Heidelberg, Germany}

\author{C.~H. Keitel}
\affiliation{Max~Planck~Institute for Nuclear Physics, Saupfercheckweg~1, D~69117 Heidelberg, Germany}

\begin{abstract}

Differences in nuclear charge distributions between isotopes lead to small changes in atomic spectra known as the field shift.
While largely proportional to the change in the mean-square nuclear radius, the field shift also contains higher-order contributions
with different dependencies on nuclear moments.
Their knowledge is required in searches for new physics using King plots,
since they contribute to deviations from King-plot linearity observed experimentally in recent studies.
We present a systematic expansion of the field-shift energies in terms of nuclear parameters and test its validity against direct numerical calculations for H-like ions. We also compute leading- and higher-order field-shift corrections for alkali-like systems from Li-like to Rb-like ions, and find that their ratios are nearly independent of the ionic charge state,  agreeing with the corresponding hydrogenic $1s$ ratios on a sub-percent level.
Motivated by these findings, we introduce an approximation in which these ratios are assumed to be independent of the electronic configuration.
It is shown that, within this approximation, higher-order field-shift corrections do not induce King-plot nonlinearities. We conclude that only the small contributions beyond this approximation can contribute to the nonlinearities observed experimentally.

\end{abstract}

\maketitle

\section{Introduction}

The isotopic shifts of atomic spectra can be measured and calculated to high accuracy.
This makes them a powerful tool for probing changes in
nuclear charge distribution along isotopic chains
and determining differences in the
mean-square nuclear charge radii \cite{aufmuth:87,fricke:04}.

The isotope shift is usually described as the sum of two principal contributions: the field shift, arising from differences
in the nuclear charge distributions of the isotopes, and the mass shift,
originating from the difference in the nuclear masses.
The field shift is mostly proportional to the change in the mean-square
nuclear charge radii, $\delta \lbr r^2\rbr$; however, it also contains
higher-order contributions
with different dependencies on nuclear moments.

These higher-order effects have recently come into focus in the context of searches for new physics based on nonlinearities in King plots. A King plot \cite{king:63,king:84} is a graphical representation of isotope shifts for at least two atomic transitions, measured across several isotope pairs. Since King plots were known to be linear to high accuracy, it was proposed \cite{delaunay:17,berengut:18} that a hypothetical new-physics boson coupling between electrons and neutrons could manifest itself as a small deviation from King-plot linearity.

Since then, several isotope-shift experiments have been carried out with the aim of constraining the new-physics coupling \cite{counts:20,hur:22,door:25,wilzewski:25}.
Although they reported nonlinearities with significances of up to
$1000\,\sigma$, none of these observations has so far been attributed to new physics.
The reason is that several "old-physics" effects also contribute to King-plot nonlinearities, in particular terms beyond
$\delta \langle r^2\rangle$ in the field shift. A reliable interpretation of the observed nonlinearities therefore requires accurate knowledge of the systematic expansion of field-shift energies in terms of nuclear parameters.

Knowledge of the higher-order field-shift effects is also required for the accurate extraction of isotopic
differences in the mean-square nuclear charge radii from experimentally measured isotope
shifts. For this purpose, the so-called Seltzer expansion
\cite{seltzer:69}
is often used
\begin{align} \label{eq:00}
\delta E_{\rm FS} = C_1 \delta \lbr r^2\rbr + C_2 \delta \lbr r^4\rbr + C_3 \delta \lbr r^6\rbr+ \ldots\,,
\end{align}
where $\delta E_{\rm FS}$ is the field-shift energy and $\delta \lbr r^n\rbr$ are differences of $n$th nuclear-charge moments.
Introduced in 1969, this expansion remains in use to this day \cite{sahoo:25}.
Yet, it is frequently overlooked that it
corresponds to the first order of
perturbation theory (see also Ref.~\cite{blundell:87}) and is therefore not exactly adequate for the
present-day accuracy requirements.
In particular, it is obvious that second-order perturbation theory will induce terms proportional to
$(\delta \langle r^2\rangle)^2$, which are of the same formal order as the $\delta\langle r^4\rangle$ contribution in Eq.~(\ref{eq:00}).
Such terms have been included into recent analyses of King-plot nonlinearities
\cite{counts:20,allehabi:21,viatkina:23,hur:22,door:25,wilzewski:25}; however, no comprehensive studies of the expansion
of the field-shift energies in terms of nuclear parameters have yet been presented.
In the present work, we aim to provide a systematically improvable expansion of the field-shift energies,
together with a robust framework for calculating the corresponding coefficients.

The remaining paper is organized as follows. We start in Sec.~\ref{sec:hlike} with considering hydrogen-like ions. We construct a systematic expansion of the field-shift energies and test its convergence against exact numerical results, taking advantage of the fact that hydrogenic ions permit calculations of essentially arbitrary precision. Next, in Sec.~\ref{sec:alkali}, we perform numerical calculations of the leading and higher-order field-shift corrections for alkali-like ions, from Li-like to Rb-like charge states. Finally, in Sec.~\ref{sec:arb}, we develop a global approximation valid for arbitrary electronic configurations and discuss its applications
 and consequences for King plots.

Throughout the paper we use the relativistic units $m=\hbar=c=1$, unless stated otherwise.
We also introduce the following notations for moments of the nuclear charge distribution and their ratio: $r_C =  \lbr r^2\rbr^{1/2}$, $r_{C4} =  \lbr r^4\rbr^{1/4}$, and $\eta = r_{C4}/r_{C}$.
We note that $\eta$ is the reciprocal of the coefficient $V_{24}$ commonly used in the literature. Since we require a separate compact notation for this quantity, we will use $\eta$ instead of $V_{24}^{-1}$.

\section{Hydrogenic ions}
\label{sec:hlike}

We consider a shift of an energy level between two isotopes, arising from differences in their nuclear charge distributions, which is commonly referred to as the field shift (FS). It is induced by a modification of the nuclear potential $V$,
\begin{align}
\delta V = V(r_C,\xi) - V(r_{C0},\xi_0),
\end{align}
where $r_C$ denotes the nuclear root-mean-square (rms) radius, $\xi$ represents additional nuclear parameters of the isotope (e.g., $r_{C4}$), and the subscript “0” refers to the reference isotope. For simplicity, the explicit dependence of the nuclear potential $V$ on the radial coordinate is omitted from the notation.

By using perturbation theory in $\delta V$ and neglecting terms beyond the second order, we may write the FS difference of the Dirac energies as
\begin{equation} \label{eq1}
    \delta E_{\mathrm{FS}} = \ \langle\psi_0|\delta V|\psi_0\rangle
    +
 \lbr \psi_0 | \delta V  \frac{1}{(E_0-H_0)^{\prime}} \delta V |\psi_0 \rbr
 \,,
\end{equation}
where $H_0$ is the Dirac Hamiltonian with $V(r_{C0},\xi_0)$, $\psi_0$ and $E_0$ are the corresponding eigenfunction and eigenvalue, respectively, and $1/(E-H)^{\prime}$ is the reduced Dirac Green function.

\subsection{Fixed nuclear shape}

Let us first consider a simplified scenario in which the isotopes are assumed to have the same fixed shape of the charge distribution.
Under the fixed shape we will understand any one-parameter model of the
nuclear charge distribution that depend only on $r_C$.
The specific choice of the model determines the assumed nuclear charge-distribution shape.

In the fixed-shape scenario,
the nuclear potential is a function of a single nuclear parameter, allowing us to expand $V(r_C)$ around $r_C = r_{C0}$ in powers of $\delta r_C^2 \equiv r_{C}^2-r_{C0}^2$. Therefore,
\begin{align} \label{eq3}
\delta V = &\  V^{\prime}_{r_{C0}}\,\delta r_C^2
% \nonumber \\ &
 + \frac{1}{2} V^{\prime\prime}_{r_{C0}}\,
    \left( \delta r_C^2 \right)^2
%  \nonumber \\ &
  + \ldots\,,
\end{align}
where the short-hand notations $V^{\prime}_{r_{C}}$ and $V^{\prime\prime}_{r_{C}}$ are defined as
\begin{align}
V^{\prime}_{r_{C}} \equiv \frac{\partial V(r_C)}{\partial (r_C^2)} =&\  \frac1{2r_C}\frac{\partial V(r_C)}{\partial r_C}\,,
 \nonumber \\
V^{\prime\prime}_{r_{C}} \equiv  \frac{\partial^2V(r_C)}{\partial^2(r_C^2)} =&\ \frac1{4r_C^2}\frac{\partial ^2V(r_C)}{\partial ^2r_C} - \frac1{4r_C^3} \frac{\partial V(r_C)}{\partial r_C}\,.
\end{align}
Inserting this expansion into Eq.~(\ref{eq1}), we obtain
\begin{align}\label{eq6}
\delta E_{\rm FS} = &\
\lbr V^{\prime}_{r_{C0}} \rbr \,\delta r_C^2
 + \lbr \frac{1}{2} V^{\prime\prime}_{r_{C0}}\rbr \,
    \left(\delta r_C^2 \right)^2
 \nonumber \\ &
 +
 \lbr V^{\prime}_{r_{C0}}  \frac{1}{(E_0-H_0)^{\prime}}  V^{\prime}_{r_{C0}}  \rbr
 \left( \delta r_C^2 \right)^2
+\ldots \,.
\end{align}
The first term here is the familiar leading-order FS contribution, whereas the remaining terms describe higher-order corrections.

It should be pointed out that a fully relativistic treatment of the finite nuclear size effect leads to appearance of non-integer powers of $r_C$ (or, equivalently, logarithms in the $Z\alpha$-expansion) \cite{shabaev:93:fns}, see also Ref.~\cite{flambaum:18,munro:22}. E.g., for $s$ states,
\begin{align}
\lbr V(r_C)+\frac{\Za}{r} \rbr \propto &\
\left(Z\alpha r_C\right)^{2\gamma}
= \left(Z\alpha r_C\right)^2
 \nonumber \\ & \times
 \left[
 1 - (\Za)^2\,\ln\left( Z\alpha r_C\right)
 + \ldots
 \right]\,,
\end{align}
where $\gamma = \sqrt{1-(Z\alpha)^2}$,
with $Z$ being the nuclear charge number and $\alpha$, the fine-structure constant.
The appearance of the logarithms $\ln r_C$ in the $Z\alpha$ expansion can be traced back to the singularity of the point-nucleus Coulomb potential, i.e., the limit $r_C=0$. For the isotope shift, however, only nuclear radii in a small vicinity of the reference-isotope radius $r_{C0}$ are relevant.
For $r_C \approx r_{C0}$, $\ln r_C$ is a smooth function that can be expanded in a well-converging Taylor series.
This justifies an expansion in integer powers of $\delta r_C^2$ when treating isotope shifts. In contrast, for the nuclear-size correction for a single isotope, one has to subtract the point-nucleus limit and thereby deal with the Coulomb singularity.
In this case one can find that the analog of the expansion (\ref{eq6}) converges very slowly.

It is instructive to examine the relative magnitude of the higher-order FS corrections in practical situations.
To this end, we performed numerical calculations of the FS energies, both within the expansion (\ref{eq6}) and by directly taking the numerical difference of the energies of the two isotopes.
Our numerical results for the ground state of hydrogen-like iron and samarium ions are summarized in Table~\ref{tab:1}. Owing to the simplicity of the one-electron systems, the calculations can be performed with essentially arbitrary numerical precision, so all reported digits are significant.
For the purposes of the present calculations, the nuclear parameters of the isotopes are assumed to be given exactly by the values listed in the table.

Our computations were performed using the dual-kinetic-balance method \cite{shabaev:04:DKB}. This method provides an accurate representation of the (discretized) spectrum of the Dirac Hamiltonian in a finite-basis space constructed from $B$-splines. Using this method, we were able to compute the matrix elements, as well as the reduced Dirac--Coulomb Green function in Eq.~(\ref{eq6}).
To avoid possible numerical round-off errors, the finite-basis method was implemented using quadruple-precision arithmetic, corresponding to approximately 32 decimal digits. This implementation has previously been used for high-precision calculations of the nuclear recoil effect \cite{yerokhin:15:recprl} and the three-photon nuclear-size effect in hydrogen \cite{pachucki:18}.
To model the nuclear charge distribution, we employed the one-parameter Fermi distribution $\rho_{\rm 1pF}$ described in Appendix~\ref{sec:app}.

Analyzing results presented in Table~\ref{tab:1}, we find very good agreement between the results obtained from the expansion (\ref{eq6}) and the direct numerical evaluation. The two higher-order FS terms in Eq.~(\ref{eq6}) reproduce about 98\% of the total higher-order contribution (given by the difference between the ``Direct'' and ``L.o.'' values). We also observe that both higher-order FS corrections are of comparable magnitude, with the second-order $V^{\prime}$ contribution being approximately twice as large as the $V^{\prime\prime}$ term.

\subsection{Varied nuclear shape}

We now would like to include effects
arising from differences in the {\em shape}
of the nuclear charge distributions between isotopes.
Describing such effects requires a model with at least two free parameters.
We here will employ the two-parameter Fermi (2pF) model, defined in Appendix \ref{sec:app}.

%%%%%%%%%%%%%%%%%%%%%%%%%%%%%%%%%%%
%
%
%
%%%%%%%%%%%%%%%%%%%%%%%%%%%%%%%%%%%
\begin{table}[t]
\caption{
Field-shift corrections within the 1pF model, for the $1s$ state of H-like Fe ($Z=26$) and Sm ($Z=62$) ions.
``L.o.'' denotes the leading-order contribution in Eq.~(\ref{eq6}),
``H.o.'' denotes the sum of the two higher-order corrections, ``Sum'' denotes
the sum of the leading and higher-order corrections, and
``Direct'' labels results of the direct numerical evaluation.
Nuclear radii are \cite{angeli:13}: $r_C(^{58,54}\mathrm{Fe}) = (3.7745, 3.6933)$~fm,
$r_C(^{154,144}\mathrm{Sm}) = (5.1053,4.9524)$~fm.
} \label{tab:1}
\begin{center}
\begin{ruledtabular}
\begin{tabular}{lw{2.6}w{2.6}}
 Term
              &  \multicolumn{1}{c}{$^{58,54}$Fe$^{25+}$ [meV]}
                &  \multicolumn{1}{c}{$^{154,144}$Sm$^{61+}$ [eV]}
\\ \hline\\[-9pt]
L.o., $V^{\prime}_{r_C}$                                          &  2.246\,350 &  0.432\,549 \\[2pt]
$V^{\prime}_{r_C} \frac1{(E-H)^{\prime}} V^{\prime}_{r_C}$  & -0.000\,672 & -0.001\,078 \\[2pt]
$\frac12 V^{\prime\prime}_{r_C}$                            & -0.000\,364 & -0.000\,511 \\[2pt]
H.o.                                                        & -0.001\,036 & -0.001\,590 \\[2pt]
Sum                                                         &  2.245\,315 &  0.430\,959 \\[2pt]
Direct                                                     &  2.245\,330 &  0.430\,994 \\
\end{tabular}
\end{ruledtabular}
\end{center}
\end{table}
%%%%%%%%%%%%%%%%%%%%%%%%%%%%%%%%%%%
%
%
%
%%%%%%%%%%%%%%%%%%%%%%%%%%%%%%%%%%%
\begin{table}[H]
\caption{
Field-shift corrections within the 2pF model with $r_C$ and $r_{C4}$ independent parameters, see Eq.~(\ref{eq8}).
Notations and nuclear radii are the same as in Table~\ref{tab:1},
further nuclear parameters are
\cite{vries:87}:
$\eta(^{58,54}\mathrm{Fe}) = (1.0789, 1.0771)$,
$\eta(^{154,144}\mathrm{Sm}) = (1.0730, 1.0650)$.
} \label{tab:2}
\begin{center}
\begin{ruledtabular}
\begin{tabular}{lw{2.6}w{2.6}}
 Term
              &  \multicolumn{1}{c}{$^{58,54}$Fe$^{25+}$ [meV]}
                &  \multicolumn{1}{c}{$^{154,144}$Sm$^{61+}$ [eV]}
\\ \hline\\[-9pt]
L.o., $V^{\prime}_{r_C}$                                               &  2.262\,403 &  0.448\,806 \\[2pt]
$V^{\prime}_{r_{C4}}$                                            & -0.019\,051 & -0.022\,153 \\[2pt]
$V^{\prime}_{r_C} \frac1{(E-H)^{\prime}} V^{\prime}_{r_C}$       & -0.001\,323 & -0.002\,167 \\[2pt]
$\frac12 V^{\prime\prime}_{r_C}$                                 & -0.000\,081 &  0.000\,293 \\[2pt]
$2\, V^{\prime}_{r_C} \frac1{(E-H)^{\prime}} V^{\prime}_{r_{C4}}$&  0.000\,981 &  0.001\,900 \\[2pt]
$V^{\prime\prime}_{r_C r_{C4}}$                                  & -0.000\,105 & -0.001\,252 \\[2pt]
$\frac12 V^{\prime\prime}_{r_{C4}}$                              &  0.000\,214 &  0.001\,128 \\[2pt]
$V^{\prime}_{r_{C4}} \frac1{(E-H)^{\prime}} V^{\prime}_{r_{C4}}$ & -0.000\,250 & -0.000\,569 \\[2pt]
H.o.                                                             & -0.019\,615 & -0.022\,821 \\[2pt]
Sum                                                              &  2.242\,787 &  0.425\,985 \\[2pt]
Direct                                                           &  2.242\,783 &  0.425\,953 \\
\end{tabular}
\end{ruledtabular}
\end{center}
\end{table}

%%%%%%%%%%%%%%%%%%%%%%%%%%%%%%%%%%%
%
%
%
%%%%%%%%%%%%%%%%%%%%%%%%%%%%%%%%%%%
\begin{table}[H]
\caption{
Field-shift corrections within the 2pF model with $r_C$ and $a$ independent parameters, see Eq.~(\ref{eq9}).
Notations and nuclear parameters are the same as in Table~\ref{tab:2}.
\label{tab:3}}
\begin{center}
\begin{ruledtabular}
\begin{tabular}{lw{2.6}w{2.6}}
 Term
              &  \multicolumn{1}{c}{$^{58,54}$Fe$^{25+}$ [meV]}
                &  \multicolumn{1}{c}{$^{154,144}$Sm$^{61+}$ [eV]}
\\ \hline\\[-9pt]
L.o., $V^{\prime}_{r_C}$                                         & 2.246\,253 &  0.432\,525 \\[2pt]
$V^{\prime}_{a}$                                                 &-0.002\,451 & -0.004\,961 \\[2pt]
$V^{\prime}_{r_C} \frac1{(E-H)^{\prime}} V^{\prime}_{r_C}$       &-0.000\,671 & -0.001\,078 \\[2pt]
$\frac12 V^{\prime\prime}_{r_C}$                                 &-0.000\,355 & -0.000\,510 \\[2pt]
$2\, V^{\prime}_{r_C} \frac1{(E-H)^{\prime}} V^{\prime}_{a}$     & 0.000\,072 &  0.000\,238 \\[2pt]
$V^{\prime\prime}_{r_C a}$                                       &-0.000\,055 & -0.000\,180 \\[2pt]
$\frac12 V^{\prime\prime}_{a}$                                   &-0.000\,019 & -0.000\,092 \\[2pt]
$V^{\prime}_{a} \frac1{(E-H)^{\prime}} V^{\prime}_{a}$           &-0.000\,004 & -0.000\,028 \\[2pt]
H.o.                                                             &-0.003\,485 & -0.006\,612 \\[2pt]
Sum                                                              & 2.242\,768 &  0.425\,914 \\[2pt]
Direct                                                           & 2.242\,783 &  0.425\,953 \\
\end{tabular}
\end{ruledtabular}
\end{center}
\end{table}

For models with more than one parameter, there is some flexibility in constructing the FS expansion, since we can chose two independent nuclear parameters differently.
Let us begin by employing the first two charge moments, $r_C$ and $r_{C4}$, as two independent nuclear-model parameters, in the spirit of Eq.~(\ref{eq:00}). Thus, $V = V(r_C, r_{C4})$. Noting that the 2pF distribution depends on $r_C^2$ and $r_{C4}^4$, we generalize the expansion (\ref{eq3}) as
\begin{align} \label{eq8}
\delta V = &\
    V^{\prime}_{r_{C0}}\,\delta r_C^2
+  V^{\prime}_{r_{C4,0}}\, \delta r_{C4}^4
 \nonumber \\ &
 + \frac{1}{2}   V^{\prime\prime}_{r_{C0}}
     \left( \delta r_C^2 \right)^2
 + \frac{1}{2}   V^{\prime\prime}_{r_{C4,0}}
     \left( \delta r_{C4}^4 \right)^2
 \nonumber \\ &
 + V^{\prime\prime}_{r_{C0}r_{C4,0}}
     \delta r_C^2\,
     \delta r_{C4}^4 + \ldots\,,
\end{align}
where
$\delta r_{C4}^4 \equiv r_{C4}^4 - r_{C4,0}^4$,
$V^{\prime}_{r_C} \equiv \partial\, V_C(r_{C},r_{C4})/(\partial\,r_{C}^2)$,
$V^{\prime}_{r_{C4}} \equiv \partial\, V_C(r_{C},r_{C4})/(\partial\,r_{C4}^4)$,
etc.
Substituting Eq.~(\ref{eq8}) into Eq.~(\ref{eq1}), we obtain the expression for the FS energies.
The resulting expression is straightforward but lengthy, and therefore is not written out explicitly.

Our numerical results for the FS contributions corresponding the expansion (\ref{eq8}) are summarized in Table~\ref{tab:2}.
Values for the ratios $\eta = r_{C4}/r_C$ for different isotopes were obtained by integrating the Fourier-Bessel expansions for the charge densities derived from electron scattering \cite{vries:87}.

Results in Table~\ref{tab:2} appear rather counterintuitive, especially for samarium.
First,
the leading-order contribution is significantly farther away from the full result than in the one-parameter case (because the derivative with respect to $r_C^2$ is now evaluated at fixed $r_{C4}$).
Furthermore, the higher-order FS corrections are much larger.
There are many of them, each with a different dependence on the nuclear parameters, and there is a significant
numerical cancellation.
Clearly, this is not an optimal situation, especially since we aim to parameterize the higher-order contribution in terms of as few combinations of nuclear parameters as possible.

This unsatisfactory situation can be attributed to the fact that $r_C$ and $r_{C4}$ do not constitute a proper choice of independent parameters.
By performing calculations with different nuclear charge-density models and nuclear radii, one finds that the values of $r_C$ and $r_{C4}$ are strongly correlated.
(In fact, in any one-parameter model,  $r_C$ and $r_{C4}$  are 100\% correlated, so it is not at all surprising.)

This correlation can be removed by introducing the ratio $\eta = r_{C4}/r_C$ \cite{yerokhin:26:rms}, suggesting that a more appropriate choice of independent variables would be $(r_C, \eta)$.
However, this would also lead to a redefinition of the leading-order contribution, which is inconvenient for our purposes.

We finally decided to use the Fermi diffuseness parameter $a$ (see Appendix \ref{sec:app}) as the second independent parameter of the charge distribution.
We recall that the one-parameter Fermi model corresponds to a fixed value of $a$.
Therefore, all derivatives with respect to $r_C$ in the two-parameter model, evaluated at fixed $a$, coincide with those in the corresponding one-parameter model.
The terms involving derivatives with respect to $a$ can therefore be identified as arising from differences in the shape of the nuclear charge distribution between the two isotopes.

So, we consider the electrostatic potential from the 2pF nuclear charge distribution with two independent nuclear parameters $r_C$ and $a$,  $V = V_{\rm 2pF}(r_C,a)$. In this case,
the expansion of $V$ around the point $(r_{C0},a_0)$ is
\begin{align} \label{eq9}
\delta V = &\
    V^{\prime}_{r_{C0}}\,\delta r_C^2
+  V^{\prime}_{a_0}\,\delta a^2
 \nonumber \\ &
 + \frac{1}{2}   V^{\prime\prime}_{r_{C0}}
     \left( \delta r_C^2  \right)^2
 + \frac{1}{2}   V^{\prime\prime}_{a_0}
     \left( \delta a^2 \right)^2
 \nonumber \\ &
 + V^{\prime\prime}_{r_{C0}a_0}
     \delta r_C^2 \,
     \delta a^2 + \ldots\,,
\end{align}
where
$\delta a^2 \equiv a^2 - a_0^2$,
$V^{\prime}_{a} \equiv \partial\, V_C(r_{C},a)/\partial(a^2)$,
etc. The corresponding expression for the FS energy is obtained after inserting the above expansion into Eq.~(\ref{eq1}).

Table~\ref{tab:3} summarizes our numerical results obtained for the 2pF model with $r_C$ and $a$ treated as independent parameters.
As expected, we find that
the leading-order contribution is nearly identical with that of Table~\ref{tab:1}; the small deviation is due to the difference in the $a$ parameters.
The higher-order FS corrections are significantly smaller than those in Table~\ref{tab:2} and numerical cancellations between them are absent.
Notably, the total higher-order FS contribution is substantially larger (by a factor of six for Sm!) than in the one-parameter case.
This difference arises from terms involving $V'_a$, which originate from change in the shape of the nuclear charge distribution between the two isotopes. This suggests that analyses of higher-order FS corrections that assume an identical nuclear charge shape may miss the dominant contribution.

Numerical results in Table~\ref{tab:3} demonstrate a clear hierarchy among the higher-order FS contributions:
the dominant correction arises from $V^{\prime}_{a}$, followed by the second-order $V^{\prime}_{r_C}$ term, and then by $V^{\prime\prime}_{r_C}$.
The sum of these three contributions accounts for about 99\% of the total higher-order FS correction.
The remaining terms are not negligible individually, but largely cancel each other.
%In the rest of this work, we will retain only the three leading higher-order FS contributions.

\subsection{Summary}

Summarizing our analysis presented so far, we write the expansion of the relativistic FS energies in a form suitable for a general atomic system, keeping the three largest higher-order corrections. The result is
\begin{align} \label{eq7}
\delta E_{\rm FS} = F^{(1)}\,\delta r_C^2
       + G_{a}\, \delta a^2
       + \left[ F^{(2)}
       + G_{r_C}\right]\, (\delta r_C^2)^2 \,,
\end{align}
where the FS constants are defined by
\begin{align} \label{eq7b}
F^{(1)} =& \  \Big< \frac{\partial V(r_C,a)}{\partial (r_C^2)} \Big>_0\,,
 \\  \label{eq7c}
G_{a} =& \  \Big< \frac{\partial V(r_C,a)}{\partial (a^2)} \Big>_0\,,
  \\ \label{eq7d}
G_{r_C}  =&\ \Big< \frac12 \frac{\partial^2 V(r_C,a)}{\partial^2 (r_C^2)} \Big>_0\,,
  \\ \label{eq7e}
F^{(2)} =&\ \Big< \frac{\partial V(r_C,a)}{\partial (r_C^2)}
  \frac1{(E-H)^{\prime}}
  \frac{\partial V(r_C,a)}{\partial (r_C^2)} \Big>_0\,,
\end{align}
where the subscript ``0'' on the matrix element indicates that it should be evaluated with the nuclear parameters of the reference isotope.
It is important to keep in mind that the derivatives with respect to $r_C$ and $a$ depend on the choice of the other independent nuclear parameter. In Eqs.~(\ref{eq7b})-(\ref{eq7e}), derivatives with respect to $r_C$ are taken at fixed $a$, while derivatives with respect to $a$ are taken at fixed $r_C$.

An appealing feature of Eq.~(\ref{eq7}) is that it explicitly separates the
effect arising from the change in the nuclear shape (the term proportional
to $\delta a^2$). However, this is also a drawback, since $\delta a^2$ is
a parameter specific to the chosen nuclear model (2pF). One can easily rewrite
$\delta a^2$ in terms of model-independent parameters by using Eq.~(\ref{eq:2pF:2}),
\begin{align} \label{eq12}
\delta a^2 \approx \frac{\partial a^2}{\partial (r_C^2)}\delta r_C^2 +
\frac{\partial a^2}{\partial (\eta^4)}\delta \eta^4  \,,
\end{align}
where derivatives are given in Appendix~\ref{sec:app}.
In this case the field-shift expansion of Eq.~(\ref{eq7}) takes the model-independent form
\begin{align} \label{eq7ext}
\delta E_{\rm FS} =&\
\left[ F^{(1)} + G_a \frac{\partial a^2}{\partial (r_C^2)}\right]\,\delta r_C^2
       + G_{a}\, \frac{\partial a^2}{\partial (\eta^4)}\,\delta \eta^4
\nonumber \\ &
       + \left[ F^{(2)}
       + G_{r_C}\right]\, (\delta r_C^2)^2 \,.
\end{align}

Alternatively, we could have expressed $\delta a^2$ in terms of $r_C$ and $r_{C4}$, which would lead to
\begin{align} \label{eq13}
\delta a^2 \approx \frac{\partial a^2}{\partial (r_C^2)}\delta r_C^2 +
\frac{\partial a^2}{\partial (r_{C4}^4)}\delta r_{C4}^4  \,.
\end{align}
Although this expression is formally correct, the strong correlation between $r_C$ and $r_{C4}$ suggests that there may be unnecessary cancellations between the two parts of it.
In fact, for the 1pF model $\delta a^2 = 0$, so the two parts must
cancel identically. For the 2pF model with  $\delta a^2 \neq 0$ the cancellation will be
not exact but still present. E.g., for
 $^{58,54}$Fe with the nuclear parameters from Table~\ref{tab:3}, Eq.~(\ref{eq13}) yields
$\delta a^2 = -0.13 + 0.15  = 0.02$~fm$^2$,
whereas Eq.~(\ref{eq12}) gives $\delta a^2 = 0.009 + 0.011  = 0.02$~fm$^2$.
This example once again illustrates that using $r_{C4}$ as an independent parameter in the FS expansion is not an optimal choice.

Comparing Eq.~(\ref{eq7}) with the FS expansion used previously in the literature \cite{counts:20,allehabi:21}, we recover their ansatz by omitting $G_{r_C}$ and expanding $\delta a^2$ in terms of $r_C$ and $r_{C4}$ according to Eq.~(\ref{eq13}). We note that the $G_{r_C}$ term typically contributes about 50\% of $F^{(2)}$ and therefore represents a significant contribution.

We now address the question of model dependence in the present treatment. Table~S1 in the Supplementary Material presents our results for the field-shift corrections obtained with the one-parameter homogeneously charged sphere (1pH) model, which should be compared with the results for the 1pF model presented in Table~\ref{tab:1}. We observe that the higher-order FS corrections do not depend significantly on the choice of nuclear model.
For the variable nuclear-shape case, the 2pF distribution reproduces the first two moments of the nuclear charge distribution, $r_C$ and $r_{C4}$. It was demonstrated in Ref.~\cite{yerokhin:26:rms} that fixing the first two nuclear moments substantially reduces the model dependence of the nuclear-size effect compared with the case in which only the first nuclear moment is fixed. We therefore expect that the present treatment based on the 2pF model
provides a reasonable description of the leading-order nuclear-shape effect.

Table~\ref{tab:4} presents our numerical values for the ratios
$G_a/F^{(1)}$, $G_{r_C}/F^{(1)}$, and $F^{(2)}/F^{(1)}$
for various H-like ions. The nuclear radii were taken from Ref.~\cite{angeli:13}, while the values of the parameter $\eta$ correspond to the 1pF model and were obtained using Eq.~(\ref{eq:2pF:0a}).
The extended version of Tab.~\ref{tab:4} is available in Supplementary Material as Tab.~S2.

It should be pointed out that Eq.~(\ref{eq7}) represents only the relativistic FS energies.
In addition, there are radiative QED finite-nuclear-size effects arising from the electron self-energy and vacuum polarization, which have been studied in dedicated works \cite{yerokhin:11:fns,yang:26}. These studies demonstrated that, in order to account for the radiative QED contribution, $\delta E_{\rm FS}$ in Eq.~(\ref{eq7})
should be multiplied by the radiative prefactor,
\begin{align} \label{eq13b}
\delta E_{\rm FS} \to \delta E_{\rm FS} \Big( 1 + x_{\rm rad}\Big)\,,
\end{align}
where $x_{\rm rad}$ is tabulated in Tab.~S2, basing on results reported in Ref.~\cite{yerokhin:11:fns}.
To the accuracy relevant for the present investigation, we can consider the fractional QED contribution $x_{\rm rad}$ to be independent of the nuclear parameters.

With the aid of Tables~\ref{tab:4} and S2, one can obtain fractional contributions of the higher-order FS corrections to isotope shifts.
As will be shown in the next Section, these fractional contributions are nearly independent of the electronic structure of the atom and can therefore be applied, to a good approximation, to many-electron atoms with the same nucleus.
Further discussion of the use of these tables is deferred to Sec.~\ref{sec:arb}.

%%%%%%%%%%%%%%%%%%%%%%%%%%%%%%%%%%%
%
%
%
%%%%%%%%%%%%%%%%%%%%%%%%%%%%%%%%%%%
\begin{table}[tb]
\caption{
Ratios of the higher-order FS constants defined in Eqs.~(\ref{eq7b})-(\ref{eq7e}) to the leading-order contribution, for the $1s$ state of various H-like ions. All moments of the nuclear charge distribution are expressed in femtometers. An extended version of this table is available in Supplementary Material.
} \label{tab:4}
\begin{center}
\begin{ruledtabular}
\begin{tabular}{lllw{1.6}w{2.5}w{2.5}}
\multicolumn{1}{c}{$Z$}
              &  \multicolumn{1}{c}{$r_C$}
              &  \multicolumn{1}{c}{$\eta$ }
                &  \multicolumn{1}{c}{$G_{a}/F^{(1)}$}
                &  \multicolumn{1}{c}{$G_{r_C}/F^{(1)}$}
                &  \multicolumn{1}{c}{$F^{(2)}/F^{(1)}$}
 \\
 & & & \multicolumn{1}{c}{} & \multicolumn{1}{c}{$\times 10^3$} & \multicolumn{1}{c}{$\times 10^3$}
\\ \hline\\[-9pt]
 10 &  3.0055 &  1.1147 &   -0.018599 &   -0.17687 &   -0.12104 \\
 20 &  3.4776 &  1.0957 &   -0.024176 &   -0.19316 &   -0.33439 \\
 30 &  3.9283 &  1.0840 &   -0.045252 &   -0.30603 &   -0.57918 \\
 40 &  4.2694 &  1.0776 &   -0.075450 &   -0.44593 &   -0.87218 \\
 50 &  4.6519 &  1.0721 &   -0.112848 &   -0.57614 &   -1.16384 \\
 60 &  4.9123 &  1.0692 &   -0.157875 &   -0.73333 &   -1.53564 \\
 70 &  5.3108 &  1.0655 &   -0.208920 &   -0.84502 &   -1.84788 \\
 80 &  5.4648 &  1.0643 &   -0.268069 &   -1.03110 &   -2.36981 \\
 90 &  5.7848 &  1.0621 &   -0.332537 &   -1.15439 &   -2.81910 \\
\end{tabular}
\end{ruledtabular}
\end{center}
\end{table}

\section{Alkali-like ions}
\label{sec:alkali}

We describe an atom with the relativistic no-pair Dirac-Coulomb-Breit (DCB) Hamiltonian $H_{\rm DCB}$ which is a sum of the zeroth-order Hamiltonian $H_0$
and the residual electron-electron interaction $V_I$, $H_{\rm DCB} = H_0 + V_I$, where
\begin{equation}
   H_0  = \ \sum_i \Big[ \balpha_i\cdot\bfp_i + \beta_i\,m + V(r_i)
   + U(r_i)\Big]\,,
\end{equation}
and
\begin{equation}
    V_I = \  \sum_{i < j} \Lambda_{++}\, I(r_{ij})\, \Lambda_{++}
  - \sum_{i} \Lambda_+ \, U(r_i)\, \Lambda_+\ .
\end{equation}
Here, $i$ and $j$ numerate the electrons,
$V$ is the nuclear binding potential,
$U$ is a screening potential,
$\Lambda_{+}$ and $\Lambda_{++}$ are the one-body and two-body
projection operators to the positive-energy part of the $H_0$ spectrum,
and $I$ is the electron-electron interaction operator given by
the sum of the Coulomb and the Breit interactions,
\begin{align}
    I(r_{ij}) = \  \frac{\alpha}{r_{ij}} -\frac{\alpha}{2r_{ij}}
   \Big[ \balpha_i\cdot\balpha_j + (\balpha_i\cdot\hat{\bfr}_{ij})(\balpha_j\cdot\hat{\bfr}_{ij}) \Big]\,,
   \label{eq:3}
\end{align}
where $\hat{{\bfr}} = {\bfr}/|\bfr|$.

Within many-body perturbation theory (MBPT), the energy of the valence state $v$ of an alkali-like atom is
represented by perturbation expansion in $V_I$,
\begin{align}
E = E^{(0)}+ E^{(1)} + E^{(2)} + \ldots\,,
\end{align}
where the individual contributions are given by \cite{blundell:87:adndt}:
\begin{align} \label{eq:mbpt}
E^{(0)} =& \ \vare_v + \sum_a \vare_a \, ,\\
E^{(1)} =& \big(V_{\rm HF}-U\big)_{vv} + \sum_a \big(\frac12 V_{\rm HF}-U\big)_{aa}\,,\\
E^{(2)} =&\
-\frac12 \sum_{abmn} \frac{I_{abmn}\, I_{mn;ab}}{\epsilon_{mn}-\epsilon_{ab}}
 \nonumber \\
 &- \sum_{am} \frac{(V_{\rm HF}-U)_{am}\, (V_{\rm HF}-U)_{ma}}{\vare_m-\vare_a}
 \nonumber \\
 &- \sum_{amn} \frac{I_{vamn}\, I_{mn;va}}{\epsilon_{mn}-\epsilon_{va}}
 %\nonumber \\
 + \sum_{abm} \frac{I_{abmv}\, I_{mv;ab}}{\epsilon_{vm}-\epsilon_{ab}}
 \nonumber \\
 &+ 2\,\sum_{am} \frac{(V_{\rm HF}-U)_{am}\, I_{mv;va}}{\vare_m-\vare_a}
 \nonumber \\
 &-\sum_{i\neq v} \frac{(V_{\rm HF}-U)_{vi}\, (V_{\rm HF}-U)_{iv}}{\vare_i-\vare_v}\,.
  \label{eq:mbpt:end}
\end{align}
The above formulas use the standard notations from Ref.~\cite{blundell:87:adndt}: the letters $a$, $b$, $c$, $\ldots$ designate occupied core orbitals;
$n$, $m$, $r$, $\ldots$ signify excited orbitals outside the core, including the
valence orbital; $i$, $j$, $k$, $\ldots$ can be either excited or occupied orbitals; the letter $v$ stands for the valence orbital. The operator
$V_{\rm HF}$ is defined by its matrix elements,
\begin{align}
\lbr i | V_{\rm HF}| j\rbr \equiv ( V_{\rm HF} )_{ij} = \sum_a I_{ai;aj} \,.
\end{align}
Furthermore, $I_{abcd} \equiv \lbr ab|I|cd\rbr$, $I_{ab;cd} \equiv I_{abcd} - I_{abdc}$. In addition, $\vare_i$ is the Dirac energy of the one-electron state $i$ and $\epsilon_{ab} \equiv \vare_a + \vare_b$.

Formulas (\ref{eq:mbpt})-(\ref{eq:mbpt:end})
contain terms that are independent of the valence orbital $v$. These terms represent the interaction between core electrons and  do not affect transition energies between the ground state and valence-excited states. Accordingly, they may be omitted when only transition energies are of interest.

In the present work, we need to evaluate the first- and second-order matrix elements of several one-body operators appearing in Eqs.~(\ref{eq7b})-(\ref{eq7e}). To this end, we employ the \textit{finite-field} approach, in which the perturbing operators are added to the DCB Hamiltonian and the corresponding derivatives of the eigenvalues with respect to the perturbation strength are evaluated numerically. More specifically, to determine the field-shift constants $F^{(1)}$ and $F^{(2)}$, we add the perturbing potential $\partial V(r_C)/\partial(r_C^2)$, multiplied by a small parameter $h$, to the nuclear potential, i.e.,
$V \to V + h\,\partial V(r_C)/\partial(r_C^2)$.
The eigenvalues $E(h)$ of the modified Hamiltonian are then calculated using the MBPT formulas outlined above. The field-shift constants are subsequently obtained as numerical derivatives of the energy with respect to $h$,
\begin{equation}\label{eq19}
F^{(1)} = \left. \frac{\partial E(h)}{\partial h}\right|_{h = 0}\,, \ \ \
F^{(2)} = \left. \frac12 \frac{\partial^2 E(h)}{\partial h^2}\right|_{h = 0}\,.
\end{equation}
The other field-shift constants, $G_a$ and $G_{r_C}$, were calculated in a similar way by adding the perturbing potentials $h\, \partial V(r_C)/\partial(a^2)$
and $h \,(1/2)\, \partial^2 V(r_C)/\partial^ 2(r_C^2)$
to the DCB Hamiltonian and differentiating with respect to $h$.

In our calculations, the potential $U$ entering the zeroth-order Hamiltonian $H_0$ is chosen to be the frozen-core Dirac-Fock potential. Within the finite-field approach, the perturbation added to $H_0$ may be also included in the self-consistent definition of the Dirac--Fock potential. It is known (see, e.g., Ref.~\cite{berengut:03}) that this is equivalent to summing an infinite class of diagrams known as the random-phase-approximation (RPA) corrections, thereby significantly improving the accuracy of the calculations.

Accordingly, for each value of the parameter $h$, we determine the Dirac--Fock potential self-consistently in the presence of the perturbation. A very high degree of convergence of the Dirac--Fock equations is essential for obtaining stable numerical derivatives in Eq.~(\ref{eq19}). To this end, we require all core- and valence-state Dirac energies to be converged to better than $5\times10^{-15}$~r.u.

The numerical derivatives were evaluated using symmetric finite-difference formulas with truncation errors of order $h^2$ and $h^4$. For the first derivative, these formulas are given by
\begin{align}\label{eq20}
f^{\prime}_{(2p)}(x) =&\ \frac{f_1-f_{-1}}{2h} \,,\nonumber \\
f^{\prime}_{(4p)}(x) =&\ \frac{-f_2+8f_1-8f_{-1}+f_{-2}}{12h} \,,
\end{align}
where $f_n=f(x+nh)$.
For the second derivative, the corresponding three- and five-point formulas were employed.
In practical calculations, the choice of the differentiation step size $h$ requires particular care. On the one hand, $h$ must be sufficiently large to avoid numerical instabilities arising from the subtraction of nearly equal function values. On the other hand, it must be small enough to ensure the validity and accuracy of the finite-difference approximations.

To determine an optimal value of $h$, we proceed as follows. We evaluate the quantity
$|f^{\prime}_{(2p)}-f^{\prime}_{(4p)}|$ as a function of $h$, gradually decreasing the step size. As long as $h$ remains sufficiently large that numerical instabilities are small, this difference scales as $h^2$. This behavior can be understood from the fact that the error of $f^{\prime}_{(4p)}$ is much smaller than that of $f^{\prime}_{(2p)}$; consequently, the difference is dominated by the error of $f^{\prime}_{(2p)}$, which is proportional to $h^2$. Once numerical instabilities become significant, however, the $h^2$ scaling is lost. We therefore take care to chose the values of $h$ such that the expected $\propto h^2$ scaling holds.

In our calculations, the summation over the Dirac spectrum in Eqs.~(\ref{eq:mbpt}) was carried out
using a finite basis set constructed from $B$-splines by the dual-kinetic-balance method \cite{shabaev:04:DKB}.
Most computations were performed in standard double-precision arithmetic (16 decimal digits). However,
the derivatives of the 2pF nuclear potential over nuclear parameters were evaluated in quadruple-precision arithmetic (32 decimal digits)
in order to suppress possible round-off errors.

Our numerical results for the leading and higher-order FS corrections in alkali-like ions are summarized in Table~\ref{tab:alkali}. The calculations were performed for transitions between the lowest-lying $p$ and $s$ states of Li-like, Na-like, K-like, and Rb-like ions. For each correction, we present results obtained within the MBPT expansion truncated at first order, $E^{(0)}+E^{(1)}$ (MBPT1), and at second order, $E^{(0)}+E^{(1)}+E^{(2)}$ (MBPT2).
We find that the difference between the MBPT1 and MBPT2 results is very small for Li-like ions, amounting to only 0.01–0.04\%.
The difference grows with increasing electron number and decreasing nuclear charge but remains below 4\% even for Rb-like charge state.

For the higher-order FS corrections, we report both their absolute values and their ratios to the leading-order contribution, $F^{(1)}$. It is remarkable that, although the corrections themselves exhibit noticeable differences between the MBPT1 and MBPT2 calculations, their ratios to $F^{(1)}$ remain unchanged to four or five significant digits. This indicates that these ratios are largely insensitive to electron–electron interactions, a feature already noted in Ref.~\cite{viatkina:23}.
The physical rationale behind this behavior is that both the leading- and higher-order FS effects are localized within the nuclear region, whereas electron correlation effects are primarily relevant at larger distances.

This observation is important because it indicates that, in calculations of higher-order FS corrections, it is advantageous to evaluate their ratios to the leading-order FS constant,
employing exactly the same method for the leading- and the higher-order corrections. Since these ratios are largely insensitive to details of the electron-electron interaction, a relatively simple treatment of correlation effects is sufficient. The main efforts should be instead directed toward ensuring numerical stability, which is a considerably more demanding task for the higher-order than for the leading-order contributions.

An even more striking observation is that the ratios of the higher-order FS corrections to $F^{(1)}$ depend only weakly on the charge state and the specific transition of the atom.
In fact, for a given isotope, these ratios are remarkably close to the corresponding hydrogenic $1s$-state values. As seen from Table~\ref{tab:alkali}, the deviation does not exceed 0.2\% for any charge state of Fe and 0.7\% for Sm.
This indicates that, at the sub-percent level of accuracy, the relative magnitude of the higher-order FS corrections can be estimated directly from the hydrogenic $1s$ values.

%%%%%%%%%%%%%%%%%%%%%%%%%%%%%%%%%%%
%
%
%
%%%%%%%%%%%%%%%%%%%%%%%%%%%%%%%%%%%
\begin{table*}[htb]
\caption{
Field-shift corrections for transitions in Li-like ($2p$-$2s$), Na-like ($3p$-$3s$),  K-like ($4p$-$4s$), and Rb-like ($5p$-$5s$) charge states of iron and samarium.
``hydr'' labels the hydrogenic $1s$ values, ``MBPT1'' and ``MBPT2'' correspond to the first-order and second-order MBPT results, respectively, ``ratio'' denotes the ratio to $F^{(1)}$. Nuclear parameters are the same as in Table~\ref{tab:3}.
} \label{tab:alkali}
\begin{center}
\begin{ruledtabular}
\begin{tabular}{lw{3.5}w{3.5}w{3.5}w{3.5}w{2.6}w{3.5}w{2.6}}
 Term      &   \multicolumn{1}{c}{$F^{(1)}$}
              & \multicolumn{2}{c}{$G_{a}$} &  \multicolumn{2}{c}{$G_{r_C}$} & \multicolumn{2}{c}{$F^{(2)}$} \\
              \cline{3-4} \cline{5-6} \cline{7-8}
              \\[-7pt]
      &\multicolumn{1}{c}{meV/fm$^2$}  &     \multicolumn{1}{c}{meV/fm$^2$}      &    \multicolumn{1}{c}{ratio}     &\multicolumn{1}{c}{$\mu$eV/fm$^4$}&    \multicolumn{1}{c}{ratio$\times10^3$}     &\multicolumn{1}{c}{$\mu$eV/fm$^4$}&    \multicolumn{1}{c}{ratio$\times10^3$}
\\ \hline\\[-7pt]
\multicolumn{4}{l}{ $Z =           26$}\\
hydr  &             &              &  -0.03308    &              &  -0.26080    &              &  -0.49281   \\
\multicolumn{4}{l}{ $2p_{1/2}$ - $2s$}\\
MBPT1 & -0.42352    &    0.014034  &  -0.03314    &    0.11064   &  -0.26124    &    0.20891   &  -0.49327    \\
MBPT2 & -0.42371    &    0.014041  &  -0.03314    &    0.11070   &  -0.26125    &    0.20900   &  -0.49327    \\
\multicolumn{4}{l}{$2p_{3/2}$ - $2s$}\\
MBPT1 & -0.42531    &    0.014086  &  -0.03312    &    0.11105   &  -0.26111    &    0.20971   &  -0.49307    \\
MBPT2 & -0.42550    &    0.014093  &  -0.03312    &    0.11111   &  -0.26112    &    0.20980   &  -0.49307    \\
\multicolumn{4}{l}{ $3p_{1/2}$ - $3s$}\\
MBPT1 & -0.07440    &    0.002465  &  -0.03313    &    0.019443  &  -0.26133    &    0.03670   &  -0.49325    \\
MBPT2 & -0.07483    &    0.002479  &  -0.03313    &    0.019557  &  -0.26135    &    0.03691   &  -0.49325    \\
\multicolumn{4}{l}{ $3p_{3/2}$ - $3s$}\\
MBPT1 & -0.07467    &    0.002473  &  -0.03312    &    0.019501  &  -0.26117    &    0.03682   &  -0.49304    \\
MBPT2 & -0.07510    &    0.002487  &  -0.03312    &    0.019616  &  -0.26118    &    0.03703   &  -0.49304    \\
\multicolumn{4}{l}{$4p_{1/2}$ - $4s$}\\
MBPT1 & -0.012945   &    0.0004289 &  -0.03313    &    0.003384  &  -0.26140    &    0.006386  &  -0.49334    \\
MBPT2 & -0.013374   &    0.0004431 &  -0.03313    &    0.003496  &  -0.26140    &    0.006598  &  -0.49332    \\
\multicolumn{4}{l}{$4p_{3/2}$ - $4s$}\\
MBPT1 & -0.012981   &    0.0004297 &  -0.03310    &    0.003391  &  -0.26126    &    0.006403  &  -0.49327    \\
MBPT2 & -0.013415   &    0.0004441 &  -0.03310    &    0.003505  &  -0.26127    &    0.006617  &  -0.49325    \\
%---------------------------------------------
%
\hline\\[-5pt]
\multicolumn{4}{l}{ $Z =           62$}\\
hydr  &             &              &  -0.16467    &              &  -0.76689    &              &   -1.6214 \\
\multicolumn{4}{l}{ $2p_{1/2}$ - $2s$}\\
MBPT1 &  -38.386    &     6.3604   &  -0.16570    &     29.620   &  -0.77165    &     62.600   &   -1.6308    \\
MBPT2 &  -38.391    &     6.3613   &  -0.16570    &     29.625   &  -0.77165    &     62.609   &   -1.6308    \\
\multicolumn{4}{l}{ $2p_{3/2}$ - $2s$}\\
MBPT1 &  -39.833    &     6.5859   &  -0.16534    &     30.671   &  -0.76998    &     64.810   &   -1.6270    \\
MBPT2 &  -39.839    &     6.5868   &  -0.16534    &     30.675   &  -0.76999    &     64.819   &   -1.6270    \\
\multicolumn{4}{l}{ $3p_{1/2}$ - $3s$}\\
MBPT1 &  -9.4369    &     1.5649   &  -0.16583    &     7.2878   &  -0.77226    &     15.396   &   -1.6315    \\
MBPT2 &  -9.4450    &     1.5663   &  -0.16583    &     7.2941   &  -0.77227    &     15.410   &   -1.6315    \\
\multicolumn{4}{l}{ $3p_{3/2}$ - $3s$}\\
MBPT1 &  -9.8195    &     1.6244   &  -0.16543    &     7.5651   &  -0.77041    &     15.979   &   -1.6273    \\
MBPT2 &  -9.8284    &     1.6259   &  -0.16543    &     7.5719   &  -0.77041    &     15.994   &   -1.6273    \\
\multicolumn{4}{l}{ $4p_{1/2}$ - $4s$}\\
MBPT1 &  -3.1609    &    0.52427   &  -0.16586    &     2.4415   &  -0.77242    &     5.1568   &   -1.6315    \\
MBPT2 &  -3.1443    &    0.52153   &  -0.16586    &     2.4288   &  -0.77243    &     5.1298   &   -1.6314    \\
\multicolumn{4}{l}{ $4p_{3/2}$ - $4s$}\\
MBPT1 &  -3.2879    &    0.54399   &  -0.16545    &     2.5334   &  -0.77053    &     5.3501   &   -1.6272    \\
MBPT2 &  -3.2674    &    0.54060   &  -0.16545    &     2.5176   &  -0.77052    &     5.3164   &   -1.6271    \\
\multicolumn{4}{l}{ $5p_{1/2}$ - $5s$}\\
MBPT1 & -0.80772    &    0.13395   &  -0.16584    &    0.62382   &  -0.77233    &     1.3175   &   -1.6312    \\
MBPT2 & -0.80437    &    0.13340   &  -0.16584    &    0.62125   &  -0.77234    &     1.3124   &   -1.6316    \\
\multicolumn{4}{l}{ $5p_{3/2}$ - $5s$}\\
MBPT1 & -0.83730    &    0.13853   &  -0.16545    &    0.64515   &  -0.77052    &     1.3624   &   -1.6271    \\
MBPT2 & -0.83452    &    0.13807   &  -0.16545    &    0.64301   &  -0.77051    &     1.3582   &   -1.6275    \\
\end{tabular}
\end{ruledtabular}
\end{center}
\end{table*}

\section{Arbitrary atomic state}
\label{sec:arb}

Based on the analysis presented in the previous section, we introduce what we refer to as the {\em global} approximation:
the ratios of the higher-order FS constants to $F^{(1)}$ are assumed to be the same for all electronic states of a given atom.
Under this assumption, the fractional contribution of the higher-order FS effects is identical to that for the $1s$ state of the corresponding hydrogenic ion.

Assuming that the ratios of the higher-order FS constants to $F^{(1)}$
are independent of the electronic structure of the atom, we
rewrite Eq.~(\ref{eq7}) as
\begin{align}
\delta E_{\rm FS} = F^{(1)}\,
\lambda_*(Z,r_C,a;r_{C0},a_0)
\,,
\end{align}
where $\lambda_*$ is the generalization of the Seltzer's moment $\lambda$,
\begin{align}\label{eq:27}
\lambda_* = \delta r_C^2 +  \frac{G_{a}}{F^{(1)}}\, \delta a^2
 + \left[ \frac{F^{(2)}}{F^{(1)}} +  \frac{G_{r_C}}{F^{(1)}}
  \right] \left( \delta r_C^2 \right)^2 \,.
\end{align}
Within the global approximation,
$\lambda_*$ depends only on nuclear parameters of the two isotopes.

It should be stressed that, despite the apparent similarity,
our moment $\lambda_*$ differs from the Seltzer moment
$\lambda$ defined in Ref.~\cite{seltzer:69} as
$\lambda = \delta \lbr r^2\rbr + (C_2/C_1)\delta \lbr r^4\rbr + (C_3/C_1)\delta \lbr r^6\rbr$.
This can be checked by looking at their numerical values. For example,
for $^{154,144}$Sm, we obtain (by taking the ratio of ``Sum'' and ``L.o.'' entries
in Table~\ref{tab:3}) $\lambda_*/\delta \lbr r^2\rbr = 0.985$,
whereas for the Seltzer moment (obtained with coefficients $C_i$ from Ref.~\cite{seltzer:69})
this ratio is $\lambda/\delta \lbr r^2\rbr = 0.94$.

The reason for this difference is that the Seltzer moment must be multiplied
by the leading-order FS contribution defined as in Table~\ref{tab:2}
(with derivative over $r_C$ evaluated at fixed $r_{C4}$).
Indeed, taking the ratio of the ``Direct'' and ``L.o.'' values from Table~\ref{tab:2},
we obtain 0.95, which is close to the Seltzer value.
This subtle aspect of the Seltzer moment is often overlooked. In particular, the leading-order FS constant $F^{(1)}$ is usually evaluated using one-parameter nuclear models (see, e.g., the recent review \cite{sahoo:25}). If such leading-order constant is combined with the Seltzer moment, the higher-order FS corrections will be significantly overestimated.

\subsection{Global approximation: a how-to guide}
We define the fractional higher-order FS contribution by
\begin{align}
f_{h.o.} = \frac{\lambda_* -  \delta r_C^2}{ \delta r_C^2}\,,
\end{align}
where $\lambda_*$ is defined by Eq.~(\ref{eq:27}). Within the global approximation,
both  $\lambda_*$ and $f_{h.o.}$ are independent of the electronic configuration and depend
only on nuclear parameters of the two isotopes. The ratios $G_a/F^{(1)}$, $G_{r_C}/F^{(1)}$,
and $F^{(2)}/F^{(1)}$ can be conveniently
taken from calculations performed in Sec.~\ref{sec:hlike} and
tabulated in Tab.~\ref{tab:4} and its extended version, Tab.~S2 in Supplementary Material.
The dependence of these ratios on nuclear parameters is weak and can be
neglected for the present purposes.

To compute $f_{h.o.}$, one also needs the differential nuclear parameters,
$\delta r_C^2$ and $\delta a^2$.
The mean-square nuclear radii $r_C^2$ are tabulated in Refs.~\cite{angeli:13,ohayon:25:radii}. We note that the uncertainties of nuclear radii
in the tabulation by Angeli \cite{angeli:13} are probably underestimated (see the discussion in
Ref.~\cite{ohayon:25:radii}) and should be used with appropriate care.

The Fermi diffuseness parameter $a$ is obtained from $r_C$ and $\eta \equiv r_{C4}/r_C$
according to Eq.~(\ref{eq:2pF:2}). The ratios of nuclear moments, $\eta$, or equivalently their reciprocals, $V_{24}$, are not well known; the first tabulation is currently in preparation \cite{ohayon:26:priv}.
One way to obtain these ratios is by integrating the Fermi-Bessel
expansion of the nuclear charge density derived from electron-scattering
data and tabulated (for a number of isotopes) in Ref.~\cite{vries:87}.
An alternative is to use results of {\em ab initio} nuclear calculations.
In particular, Ref.~\cite{miyagi:26} reported results for $^{208}$Pb,
Ref.~\cite{hur:22} presented calculations for several isotopes of Yb, and Ref.~\cite{he:26}
for $^{26}$Mg. The {\em ab initio} values
are often in remarkable agreement
with those obtained from electron-scattering data.
Nevertheless, the uncertainties associated with the $\eta$ values are not yet well understood, and the resulting uncertainty in the difference $\delta a^2$ may therefore be substantial and and not reliably quantified.

In order to help the reader to estimate the uncertainty of the global approximation,
Table~\ref{tab:comp} presents a comparison of the ratio $G_a/F^{(1)}$ evaluated for
the $1s$ state of H-like ions and for several $np_{1/2}$-$ns$ transitions of alkali-like ions
over a range of nuclear charge numbers $Z$. We observe that the deviation between
the ratios is very small, gradually increasing from 0.06\% for $Z=10$ to $1.5\%$ for $Z=90$.
Very similar behaviour was found also
for the other ratios, $G_{r_C}/F^{(1)}$ and $F^{(2)}/F^{(1)}$,
and for different transitions, in particular, $3d_{3/2}$-$4s$ and $3d_{5/2}$-$4s$
transitions in K-like ions.
We thus suggest to use the
deviations listed in Table~\ref{tab:comp} for estimating the uncertainty
of the global approximation in most cases.

We identified only one class of transitions for which the global approximation has
a larger uncertainty, namely, the fine-structure transitions, i.e.,
transitions between the states that differ only by the value of the total angular momentum.
For example, Table~\ref{tab:alkali} shows that
the deviation of the FS ratios for the $np_{3/2}$--$np_{1/2}$ intervals from hydrogenic values
is about 12\% for $Z = 26$ and 5\% for $Z=62$.
We therefore conclude that the uncertainty of the global approximation
for the fine-structure transitions is on the level of 10\%.
This effect is probably attributable to the strong cancellations occurring in
the fine-structure intervals and to the fact that these intervals arise entirely from relativistic effects.

Furthermore, there exist electronic configurations for which the approximation is
not applicable at all. The physical rationale behind the global approximation is
that the typical atomic wave function inside the nuclear region behaves similarly
to the hydrogenic $s$-state wave function, with the difference arising mostly through the
normalization constant, which cancels in the ratios.
Even when the valence orbital has high angular momentum, it remains
coupled to the core $s$ electrons through the electron-electron interaction.
Clearly, this reasoning fails if the electronic configuration contains
no $s$ electrons, in which case the global approximation is not
applicable. Examples of such configurations are the hydrogenic states with $l>1$,
doubly-excited $(nl,n'l')$ states of helium-like ions, etc.

%%%%%%%%%%%%%%%%%%%%%%%%%%%%%%%%%%%
%
%
%
%%%%%%%%%%%%%%%%%%%%%%%%%%%%%%%%%%%

\begin{table*}[t]
\caption{
Comparison of ratios $G_{a}/F^{(1)}$ for the $1s$ state of H-like ions and the $np_{1/2}$-$ns$ transitions in alkali-like ions. Nuclear parameters and model the same as in  Table~\ref{tab:4}. ``Diff.'' is the deviation from the corresponding H-like value.
} \label{tab:comp}
\begin{center}
\begin{ruledtabular}
\begin{tabular}{lw{3.6}w{3.6}w{2.4}w{3.6}w{2.4}w{3.6}w{2.4}w{3.6}w{2.4}}
%\hline\\[-7pt]
 \multicolumn{1}{c}{$Z$}
              &  \multicolumn{1}{c}{H-like, $1s$}
                &  \multicolumn{2}{c}{Li-like, $ 2p_{1/2}$-$2s$}
                &  \multicolumn{2}{c}{Na-like, $ 3p_{1/2}$-$3s$}
                &  \multicolumn{2}{c}{K-like, $ 4p_{1/2}$-$4s$}
                &  \multicolumn{2}{c}{Rb-like, $ 5p_{1/2}$-$5s$}
\\
              & & & \multicolumn{1}{c}{Diff., \%}
& & \multicolumn{1}{c}{Diff., \%}
& & \multicolumn{1}{c}{Diff., \%}
& & \multicolumn{1}{c}{Diff., \%}
\\
\hline\\[-9pt]
10 & -0.01860 & -0.01861 & 0.06  \\
20 & -0.02418 & -0.02420 & 0.10  & -0.0242 & 0.13  \\
30 & -0.04525 & -0.04535 & 0.21  & -0.0454 & 0.23 & -0.0453 & 0.18 \\
40 & -0.07545 & -0.07569 & 0.32  & -0.0757 & 0.35 & -0.0757 & 0.35 & -0.0757 & 0.37 \\
50 & -0.11285 & -0.11336 & 0.45  & -0.1134 & 0.50 & -0.1134 & 0.51 & -0.1134 & 0.49 \\
60 & -0.15788 & -0.15881 & 0.60  & -0.1589 & 0.67 & -0.1590 & 0.69 & -0.1589 & 0.68 \\
70 & -0.20892 & -0.21056 & 0.78  & -0.2108 & 0.89 & -0.2108 & 0.92 & -0.2108 & 0.91 \\
80 & -0.26807 & -0.27067 & 0.97  & -0.2711 & 1.11 & -0.2712 & 1.15 & -0.2711 & 1.15 \\
90 & -0.33254 & -0.33664 & 1.23  & -0.3373 & 1.42 & -0.3374 & 1.47 & -0.3374 & 1.47 \\
%\hline
%
\end{tabular}
\end{ruledtabular}
\end{center}
\end{table*}

%%%%%%%%%%%%%%%%%%%%%%%%%%%%%%%%%%%

\subsection{Calcium}

We now turn to comparing our results with previous calculations available in the literature.
First, we consider the isotope shift in Ca$^+$.
In Ref.~\cite{viatkina:23}, higher-order FS corrections were computed
for isotope shifts of the $4p_{j}$-$4s$ and $3d_{j}$-$4s$
transitions in Ca$^+$. Specifically, for the $^{44,40}$Ca isotope shift
Ref.~\cite{viatkina:23} reported the fractional higher-order contribution
of $ -1.48\times 10^{-4}$ for all four transitions, assuming nuclear radii from Ref.~\cite{angeli:13} and the same
shape for both isotopes.
From the data in Table~\ref{tab:4}, we
immediately obtain (with $\delta a^2=0$) $ f_{h.o.}=-1.487\times 10^{-4}$,
in perfect agreement with Ref.~\cite{viatkina:23}.

However, if we take into account the difference of nuclear shapes of the two isotopes,
the results change significantly. Using the values of the $\eta$ parameter obtained from nuclear density functional theory (DFT) calculations \cite{naito:21},
\begin{align}\label{eq:30}
\eta^{-1}\left( ^{40}\mathrm{Ca}\right) = 0.913\,,\ \
\eta^{-1}\left( ^{44}\mathrm{Ca}\right) = 0.916\,,
\end{align}
together with the rms charge radii from Ref.~\cite{angeli:13}, we obtain $\delta a^2=-0.012~\mathrm{fm}^2$. Using the corresponding value of $G_a$ from Table~\ref{tab:4}, we find a fractional contribution of the nuclear-shape effect,
$f_{h.o.}(G_a) = 0.0010$,
which is seven times larger than the value obtained under the assumption of the fixed nuclear shape.
It should be noted that the uncertainty of the nuclear-theory values of $\eta$ in Eq.~(\ref{eq:30}) is presently unclear. Consequently, the estimated magnitude for the nuclear-shape effect should be interpreted with appropriate caution.

Higher-order FS contributions were also investigated recently by Kayal et al. \cite{katyal:25}, who obtained very different results. Their FS constants $G^{(2)}$, $F^{(2)}$, and $F^{(1)}$ should correspond directly to our $G_{r_C}$, $F^{(2)}$, and $F^{(1)}$, respectively. For the $3d_{3/2}$--$4s$ transition, they obtained (in fm$^{-2}$) $G^{(2)}/F^{(1)}=-0.0096$ and $F^{(2)}/F^{(1)}=6\times10^{-7}$, whereas our corresponding values are $-0.00019$ and $-0.00033$, respectively (see Table~\ref{tab:4}).
Since our H-like results have been cross-checked against direct numerical calculations, and the ratios for Ca$^+$ should be close to the hydrogenic values, we conclude that the results reported in Ref.~\cite{katyal:25} are likely erroneous.

\subsection{Ytterbium}

The isotope shifts of Yb and Yb$^+$ were studied in Refs.~\cite{counts:20,hur:22,door:25} in the context of King-plot nonlinearities. Ref.~\cite{hur:22} also computed the nuclear charge moments of Yb isotopes
within nuclear DFT approach, yielding values for the
corresponding $\eta$ parameters. Specifically, from Table~S6 of that work we obtain, by averaging over four different DFT functionals,
\begin{align}
\eta\left(^{176}\mathrm{Yb}\right) = 1.0684\,(4)\,, \\
\eta\left(^{168}\mathrm{Yb}\right) = 1.0701\,(4)\,,
\end{align}
where uncertainties are the standard deviation of the four values.
The remarkable stability of the DFT values of $\eta$ parameters is noteworthy,
especially given that the corresponding results for $r_C$ and $r_{C4}$ exhibit
variations on a percent level.

We now analyze the results of Ref.~\cite{hur:22} for higher-order FS contributions.
In their formulation, there are two higher-order FS correction, whose fractional contributions are given by
\begin{align} \label{eq16}
\Delta_1 = \frac{G^{(2)}(\delta r_C^2)^2 }{F^{(1)}\delta r_C^2}\,, \ \
\Delta_2 = \frac{G^{(4)}\delta r_{C4}^4}{F^{(1)}\delta r_C^2}\,.
\end{align}
Using the GRASP results from Table S7 of Ref.~\cite{hur:22} and nuclear radii from Ref.~\cite{angeli:13}, we obtain for the $^2S_{1/2}$–$^2D_{5/2}$ transition in $^{176,168}$Yb$^+$ the following values: $\Delta_1 = -0.00149$
and $\Delta_2 = -0.058$. Note a very large (6\%!) contribution of the $\delta r_{C4}^4$ term. However, as we discussed previously, most of this contribution should cancel out with the $\delta r_C^2$ term.

We now will use our hydrogenic data from Table~\ref{tab:4}  to obtain our predictions for fractional higher-order FS corrections in Yb$^+$.
First, assuming identical nuclear shapes for both isotopes ($\delta a^2 = 0$), we obtain $f_{h.o.}(F^{(2)},G_{r_C}) = -0.00146$, which is in good agreement with $\Delta_1$ of Ref.~\cite{hur:22}. Calculating separately the nuclear shape $(\propto \delta a^2)$ contribution, we obtain $f_{h.o.}(G_a) = 0.00573$. We observe that
the nuclear shape correction is much smaller than $\Delta_2 \propto \delta r_{C4}^4$, as expected.
In order to compare with $\Delta_2$, we separate out $\delta r_{C4}^4$ from $\delta a^2$ according to Eq.~(\ref{eq13}) and obtain
\begin{align}
\Delta_2^* = \frac{G_a \frac{\displaystyle \partial a^2}{\displaystyle \partial(r_{C4}^4)} \delta r_{C4}^4}{F^{(1)}\delta r_C^2} = -0.047\,,
\end{align}
which is of the same sign and comparable magnitude as in Ref.~\cite{hur:22}.

\subsection{Consequences for the King plot}

The King plot \cite{king:84} is a powerful way to analyze the experimentally measured
isotope shifts. It is based on the standard representation of
the isotope shift of an energy level (or transition energy)
\begin{align} \label{eq:KP}
\delta E_{\rm IS} = K^{(1)}\, \mu + F^{(1)}\, \delta r_C^2\,,
\end{align}
where $K^{(1)}$ and $F^{(1)}$ are the leading-order mass-shift and field-shift constants, respectively, $\mu = m/M - m/M_0$, and higher-order effects are neglected.
The idea behind the King plot is based on the observation that  in the above expression the electronic and nuclear variables are separated. Specifically, the isotope-shift constants $K^{(1)}$ and $F^{(1)}$ depend on the electronic state (or transition) of the atom but not on the isotope, whereas $\mu$ and $\delta r_C^2$ are nuclear parameters that depend on the isotope but not on the electronic state.
By considering isotope shifts for two different transitions, one can eliminate the poorly known parameter $\delta r_C^2$ from the system of two equations. This leads to a linear relationship between the reduced frequencies of the two isotope shifts, known as the King plot.

An important observation is that introducing the higher-order FS corrections within the global approximation into Eq.~(\ref{eq:KP}) simply replaces $\delta r_C^2$ with $\lambda_*$. Since $\lambda_*$ is likewise independent of the electronic transition, the separation between electronic and nuclear variables remains intact. Consequently, $\lambda_*$ will be eliminated by the same procedure as in the standard King-plot analysis. We therefore conclude that, within the global approximation, higher-order FS corrections do {\em not} contribute to King-plot nonlinearities.

This means that the nonlinearities arise solely from the small parts of the higher-order FS corrections that are {\em beyond} the global approximation.
This implies that any study of such nonlinearities requires the higher-order corrections to be computed with very high accuracy, at about $10^{-4}$ level.
Such precision is generally beyond the reach of direct calculations in many-electron atoms.
For example, in Yb$^+$, the higher-order FS contributions obtained in Refs.~\cite{counts:20,hur:22} by two different methods differ by 10\% and often more.
%Our analysis in Sec.~\ref{sec:alkali} demonstrates that substantially higher numerical accuracy can be achieved by calculating the {\em ratios} of the higher-order FS corrections to $F^{(1)}$ rather than the corrections themselves.

It is important that all field-shift contributions in Eq.~(\ref{eq7}) are explicitly factorized into terms depending solely on the electronic or nuclear degrees of freedom. This factorization makes it possible to eliminate them in the analysis of experimental data using various generalized versions of the King plot \cite{berengut:20:kingplot}.

It is worth noting that there are other higher-order isotope-shift corrections that likewise do not contribute to King-plot nonlinearities at leading order. For example, the leading nuclear-mass correction to the FS simply multiplies $F^{(1)}$ by a reduced-mass factor, see, e.g., Ref.~\cite{yerokhin:20:kingsplot}. As follows from Eq.~(\ref{eq:KP}),
this factor can be absorbed into $r_C^2$, thus slightly modifying
$\delta r_C^2$ and preserving the separation between electronic and nuclear variables. Consequently, it does not contribute to the nonlinearity of the King plot.

Another example is nuclear polarization. Similarly to the FS corrections, its dominant contribution originates from the nuclear region. As a result, the ratio of the nuclear-polarization contribution to $F^{(1)}$ is largely independent of the electronic structure, as confirmed by the numerical calculations of Ref.~\cite{viatkina:23}. We therefore conclude that only a small non-contact component of the nuclear-polarization contribution should generate King-plot nonlinearities.

In Table~\ref{tab:IS}, we compare estimates of various individual isotope-shift contributions for two experimentally relevant cases: the $^{44,40}$Ca$^+$ isotope shift of the $3d_{5/2}$--$4s$ transition and the $^{176,178}$Yb$^+$ isotope shift of the $^2S_{1/2}$--$^2D_{5/2}$ transition. Some of the data are taken from previous studies \cite{viatkina:23,hur:22}, as indicated in the table. The nuclear-polarization contribution for Yb$^+$ was estimated in the present work using the approach described in Ref.~\cite{viatkina:23} and the hydrogenic approximation for the ratio of the nuclear-polarization contribution to the leading-order FS contribution. The second-order mass-shift (MS) contribution for Yb$^+$ was estimated by multiplying the leading-order MS by a factor of $2(m/M+m/M_0)$. This factor follows from the parametric dependence of the second-order MS, with 2 being the conservative factor.

We observe that nuclear-polarization and nuclear-shape effects are important even for relatively light systems such as calcium, with the second-order MS contribution being of comparable magnitude. For heavier systems such as ytterbium, however, the second-order MS contribution becomes practically negligible, as expected from its suppression by the second power of the electron-to-nucleus mass ratio $m/M$.
For Yb, the nuclear-shape effect is the dominant higher-order contribution, reflecting the strong deformation of Yb nuclei. Nevertheless, the nuclear-polarization and second-order FS contributions are comparable in magnitude to the shape effect and need to be taken into account.

%%%%%%%%%%%%%%%%%%%%%%%%%%%%%%%%%%%
%
%
%
%%%%%%%%%%%%%%%%%%%%%%%%%%%%%%%%%%%
\begin{table}[t]
\caption{
Individual isotope-shift contributions in Ca$^+$ and Yb$^+$.
``FS, shape'', ``FS, h.o.'', and ``FS, sec.'' correspond to
the field-shift corrections given by Eqs.~(\ref{eq7c})-(\ref{eq7e}),
respectively, ``MS'' stands for the mass shift.
\label{tab:IS}
}
\begin{ruledtabular}
\begin{tabular}{lw{10.10}w{10.10}}
                &  \multicolumn{1}{c}{$^{44,40}$Ca$^+$}
                &  \multicolumn{1}{c}{$^{176,178}$Yb$^+$}
\\
                &  \multicolumn{1}{c}{$3d_{5/2}$--$\,4s$}
                &  \multicolumn{1}{c}{$^2S_{1/2}$--$\,^2D_{5/2}$}
\\[2pt]
                &  \multicolumn{1}{c}{[kHz]}
                &  \multicolumn{1}{c}{[MHz]}
\\[2pt]
\hline\\[-5pt]
Leading-order FS $^*$    & \multicolumn{1}{c}{ $-11 \times 10^4$} &    \multicolumn{1}{c}{ $-8613$}  \\[2pt]
Leading-order MS  $^*$   & \multicolumn{1}{c}{ $557 \times 10^4$}   &    \multicolumn{1}{c}{ $\ \ \ 455$}  \\[2pt]
Nuclear polarization $^\dag$ & \multicolumn{1}{c}{ \ \ 263}   &    \multicolumn{1}{c}{ \ \ $-11$ }  \\[2pt]
FS, shape    & \multicolumn{1}{c}{  $-110$} &     \multicolumn{1}{c}{ \ \ $-49$ }\\[2pt]
FS, sec.             & \multicolumn{1}{c}{ \ \ \ 10}  &   \multicolumn{1}{c}{ $\ \ \ \ \ \ 9$ }  \\[2pt]
FS, h.o.             & \multicolumn{1}{c}{ \ \ \ \  6}  & \multicolumn{1}{c}{ $\ \ \ \ \ \, 4$ } \\[2pt]
MS, sec.         $^\dag$    & \multicolumn{1}{c}{  $-69$}       & \multicolumn{1}{c}{ $\ \ \ \pm 0.01$ }    \\[2pt]
\end{tabular}
\end{ruledtabular}
$^*$ from Ref.~\cite{viatkina:23} for Ca$^+$ and Ref.~\cite{hur:22} (GRASP) for Yb$^+$,\\
$^\dag$  from Ref.~\cite{viatkina:23} for Ca$^+$ and this work for  Yb$^+$.
\end{table}

\section{Conclusions}

We presented a systematically improvable expansion of the field-shift energies in terms of nuclear parameters, including contributions arising from second-order perturbation theory. The validity and convergence of this expansion were verified through comparison with direct numerical calculations for hydrogen-like ions. Within our formulation, effects associated with changes in the shape of the nuclear charge distribution are cleanly separated from those arising from variations in the nuclear radius. Furthermore, our approach avoids spurious cancellations resulting from the strong correlation between
the contribution proportional to
 $\delta \langle r^2\rangle$ and $\delta \langle r^4\rangle$.

We performed numerical calculations of the leading- and higher-order field-shift corrections for several alkali-like electronic configurations, ranging from Li-like to Rb-like ions. We found that, although the corrections themselves depend strongly on the charge state, their ratios to the leading-order field-shift  remain nearly constant and are very close to the corresponding ratios for the $1s$ state of the respective hydrogenic ion. This observation indicates that these ratios are largely insensitive to the electronic structure, implying that a relatively simple treatment of electronic correlations is sufficient for their accurate evaluation.

Our analysis of alkali-like ions enabled us to introduce the  global approximation, which assumes that the fractional field-shift contributions are the same for all charge states of a given atom. This approximation is expected to be valid at the sub-percent level for most atomic transitions. Using the global approximation, we were able to analyze recent calculations for Yb$^+$ and Ca$^+$ ions.
It has been demonstrated that differences in the shape of the nuclear charge distribution are often the dominant contribution to the field shift beyond the $\delta\langle r^2\rangle$ term.

We also demonstrated that, within the global approximation, higher-order field-shift corrections do not contribute to King-plot nonlinearities. Such nonlinearities can arise only from small state-dependent components beyond the global approximation. Consequently, investigations of King-plot nonlinearities require calculations of the higher-order field-shift corrections with exceptionally high numerical accuracy.

\section*{Acknowledgement}

We are grateful to Ben Ohayon for providing us his data  before publication and many fruitful discussions.

\appendix

\section{Fermi distribution}
\label{sec:app}

The standard two-parameter Fermi distribution model for the nuclear charge density is given by
\begin{align}\label{eq:2pF}
\rho_{\rm F}(c,a;r) = \frac{\rho_{0}}{1 + \exp[(r-c)/a]}\,,
\end{align}
where the normalization prefactor and moments of the distribution are expressed in terms of the
 polylogarithm function  ${\rm Li}_n(x)$,
\begin{align}\label{eq:2pF:0}
\rho_{0}^{-1} =&\  -8\pi a^3\, {\rm Li}_3\big(-e^{c/a}\big) \,, \\
\lbr r^n \rbr =&\ \frac{a^n(n+2)!}{2}\,
 \frac{{\rm Li}_{n+3}\big( -e^{c/a}\big) }
{{\rm Li}_3\big(-e^{c/a}\big)}\,.
\label{eq:2pF:0a}
\end{align}

If the diffuseness parameter $a$ is fixed by the standard choice $a = a_0 = 2.3/(4\ln3)$~fm,
one obtains the one-parameter Fermi (1pF) model, which depends only on the parameter $c$, or,
equivalently, the rms radius $r_C$,
$\rho_{\rm 1pF} \equiv \rho_{\rm 1pF}(r_C;r)$. The parameter $c$ is  obtained from $r_C$ by \cite{gustavsson:98}
\begin{align}\label{eq:2pF:1}
c^2 = \frac53\, r_C^2 -\frac73\, \pi^2 a^2\,.
\end{align}
We note that the above formula is approximate; however, it holds with  good accuracy and, to our purposes, it will be sufficient to treat it as exact.

If $a$ is considered as a free parameter, the Fermi model can reproduce two moments of the nuclear charge distribution, $r_C$ and $r_{C4}$. We will consider $r_C$ and $a$ as independent parameters and write $\rho_{\rm 2pF} \equiv \rho_{\rm 2pF}(r_C,a;r)$.
The parameter $a$ is connected with $r_C$ and $r_{C4}$ by \cite{gustavsson:98}
\begin{align}\label{eq:2pF:2}
a^2 = \frac{r_C^2}{8\pi^2 }
\left[\sqrt{84 \eta^4-75 }-5 \right]\,,
\end{align}
where $\eta = r_{C4}/r_C$.
Again, this formula is only approximate, but to our purposes it will be sufficient to treat it as exact.
We verified that the uncertainties in the determination of $c$ and $a$ from Eqs.~(\ref{eq:2pF:1}) and (\ref{eq:2pF:2}) are less than $1\times10^{-5}$~fm for $c$ and $1\times10^{-6}$~fm for $a$, respectively, which is much smaller than the current experimental precision in the determination of nuclear parameters.
Moreover, in the context of the present investigation, we can regard these formulas as the {\it definition} of $\rho_{\rm 2pF}(r_C,\eta)$. The resulting tiny difference with the standardly assumed 2pF distribution will have no effect on the conclusions of the present work.

In the present investigation we need also derivatives of $a$ over nuclear parameters, which are given by
\begin{align}\label{eq:2pF:2b}
\frac{\partial a^2}{\partial r_C^2} = &\ \frac{a^2}{r_C^2}\,,\\
\frac{\partial a^2}{\partial \eta^4} = &\ \frac{21 r_C^2}{4 \pi^2 \sqrt{84 \eta^4-75}}\,,\\
\frac{\partial a^2}{\partial r_{C4}^4} = &\ \frac{21 }{4 \pi^2 r_C^2 \sqrt{84 \eta^4-75}}\,. \label{eq:2pF:2c}
\end{align}

For completeness, we give here known formulas for the electrostatic potential induced by the Fermi distribution \cite{gustavsson:98}. The potential for $r<c$ is given by
\begin{align}\label{eq:2pF:3}
V_{r<c}(r) = &\ -\frac{N}{r}\bigg[
 \frac{3r}{2c} - \frac{r^3}{2c^3}+ \frac{r\pi^2a^2}{2c^3}
  - \frac{3ra^2}{c^3}S_2\left(\frac{r-c}{a}\right)
  \nonumber \\ &
  + \frac{6a^3}{c^3}S_3\left(\frac{r-c}{a}\right)
  - \frac{6a^3}{c^3}S_3\left(-\frac{c}{a}\right)
  \bigg]\,,
\end{align}
whereas for $r>c$, it is
\begin{align}\label{eq:2pF:4}
V_{r>c}(r) = &\ -\frac{N}{r}\bigg[
 1 + \frac{\pi^2a^2}{c^2}
  + \frac{3ra^2}{c^3}S_2\left(-\frac{r-c}{a}\right)
  \nonumber \\ &
  + \frac{6a^3}{c^3}S_3\left(-\frac{r-c}{a}\right)
  - \frac{6a^3}{c^3}S_3\left(-\frac{c}{a}\right)
  \bigg]\,,
\end{align}
where $S_k$ and $N$ are given by
\begin{align}
S_k(x) = &\ \sum_{n=1}^{\infty}(-1)^n \frac{e^{xn}}{n^k}\,,
 \nonumber \\
N = &\ \frac{Z\alpha}{
\displaystyle
1 + \frac{\pi^2a^2}{c^2} - \frac{6a^3}{c^3}S_3\left(-\frac{c}{a}\right)} \,.
\end{align}

%\bibliography{bibliography}
%\bibliographystyle{c:/-a-/papers/bibtex/phaip30}
%\bibliography{c:/-a-/papers/bibtex/hfst}

\end{document}